# PATTERN OF INTERNET USAGE IN CYBER CAFÉS IN MANILA: AN EXPLORATORY STUDY

Rex P. Bringula
University of the East-Manila
2219 C.M. Recto Avenue, Sampaloc, Manila, Philippines
rex_bringula@yahoo.com

Jenmart Bonifacio
University of the East-Manila
2219 C.M. Recto Avenue, Sampaloc, Manila, Philippines
jenmart_bonifacio@yahoo.com

Ana Natanauan
University of the East-Manila
2219 C.M. Recto Avenue, Sampaloc, Manila, Philippines
anacn_26@yahoo.com

Mikael Manuel
University of the East-Manila
2219 C.M. Recto Avenue, Sampaloc, Manila, Philippines
mikael.manuel@yahoo.com

Katrina Panganiban
University of the East-Manila
2219 C.M. Recto Avenue, Sampaloc, Manila, Philippines
panganiban_katrinalyn@yahoo.com

## ABSTRACT

This study determined the profile and pattern of Internet usage of respondents in cyber cafés in Manila. The study employed an exploratory-descriptive design in which a validated descriptive-survey form was used as the research instrument. Forty-seven cyber cafés in 14 districts in the City of Manila were randomly selected. There were 545 respondents. It was found that most of the respondents were Manila settlers (f = 368, 70%), students (f = 382, 73%), pursuing or had attained a college degree (f = 374, 72%), male (f = 356, 68%), young (19 and below) (f = 314, 60%), Roman Catholic (f = 423, 81%), single (f = 470, 90%), had a computer at home (f = 269, 51%), belonged to the middle-income class (f = 334, 64%), and used the Internet in the afternoon (f = 274, 50.3%, $\chi 2 = 113.98$, $DF = 2$, $p < 0.01$) once to twice a week (f = 193, 36.9%, $\chi 2 = 90.04$, $DF = 3$, $p < 0.01$). Frequency of visit of Internet users was not equally distributed during the week and Internet users showed the tendency to visit cyber cafés at a particular time of the day when grouped according to profile. The first hypothesis stated that frequency of visiting a cyber café would not be equally distributed during the week when grouped according to profile was accepted. The second hypothesis, which stated that respondents would not show a tendency to use the internet in a cyber café at a particular time of the day was rejected. The study also discusses the limitations and implications of the findings.

## INTRODUCTION

Few studies have been conducted to determine the pattern of Internet usage in cyber cafés either in foreign countries or in a Philippine setting. This posed a large gap in the existing body of literature of Internet users in public venues, such as cyber cafés. Thus, this study served to address this gap.

This exploratory-descriptive study aimed to determine the pattern of internet usage of cyber café users in Manila. Specifically, it sought answers to the following questions. (1) What is the profile of the cyber café users in terms of place of residence, age, sex, civil status, educational background, occupation, religious affiliation, personal computer ownership, and monthly family income? (2) What is the internet usage of respondents in cyber cafés in terms of hours spent per visit, visiting time of day, and frequency of visit within a week? (3) Is the frequency of visiting equally distributed within a week when grouped according to profile? and (4) Do respondents show a tendency to use the Internet in a cyber café at any particular time of the day when grouped according to profile?

Pattern of Internet usage was measured in terms of hours spent per visit and visiting time of the day (Furuholt, Wahid, & Kristiansen, 2005; Adomi, Omodeko, & Otolo, 2004; GaWC, 2008), and frequency of visit within the week (Adebayo, 2008). Cross-tabulation was employed to the profile of respondents against their hours spent per visit, visiting time of the day, and frequency of visit within the week.

Moreover, to determine the profile of respondents, the following constructs were adopted: place of residence (Adomi, Okiy, & Ruteyan, 2003; Olatokun, 2009), age (Furuholt et al., 2005; Adomi, 2007; Alao & Folorunsho, 2007; Ishii & Wu, 2006; Olatokun, 2009), sex (Haseloff, 2005; Furuholt et al., 2005; Adomi et al., 2004; Adomi, 2007 ; Adomi et al., 2003; Olatokun, 2009), employment (Furuholt et al., 2005; Adomi et al., 2004; Ishii & Wu, 2006), educational background (Haseloff, 2005; Furuholt et al., 2010; Adomi et al., 2004; Ishii & Wu, 2006; Adebayo, 2008; Olatokun, 2009), religious affiliation (Williams, Yee, & Caplan, 2008; Griffiths, Davies, & Chappell, 2004), civil status (Furuholt et al., 2005; Wahid, Furuholt, & Kristiansen, 2004; Alao & Folorunsho, 2007; Adebayo, 2008), personal computer ownership (Gurol & Sevindik, 2006), and income (Furuholt et al., 2005; Ishii & Wu, 2006; Adebayo, 2008; Olatokun, 2009). These constructs were adopted because most literature has used these in determining the profile of respondents.

The following null hypotheses were tested. (1) Frequency of visit of the respondents would not be equally distributed within the week when grouped according to profile, and (2) Respondents would not show a tendency to use the internet in a cyber café at a particular time of the day when grouped according to profile.

## METHODOLOGY

A content-validated and pre-tested survey form was used as the research instrument. The first part of the survey form gathered information on the profile of the respondents and included place of residence, age, sex, occupation, religious affiliation, civil status, computer ownership, monthly family income, and educational background. The second part gathered internet usage information of the respondents in a cyber café, such as hours spent per visit, visiting time during the day, and frequency of visit within the week.

The City of Manila (a Beta + global city) (2008), with a population of 1,660,714 (National Statistics Office, 2008), was chosen as the research locale of the study since it is a Beta+ global city. The computed minimum sample size was 400. This was equally divided among the 14 districts of Manila (30 survey forms per district).

To accommodate low return rates, 45 survey forms were distributed in each district of Manila. The survey forms were distributed evenly in cyber cafés in the morning (between 7 am to 11:59 am), afternoon (between 1 pm to 6 pm), and evening (between 6:01 pm to 9 pm). Of 630 survey forms distributed to 47 cyber cafés in Manila, 545 forms were retrieved and all were used in the analysis.

The respondents of the study were Internet users in a cyber café and were chosen regardless of age, sex, religious affiliation, etc. Cyber cafes were selected through a random walk method (Haseloff, 2005) that covered the entire city of Manila. Whenever the researcher saw a cyber café, survey forms were distributed in that café. If more than one café was found on the same street, the cyber café was chosen alternately.

The statistical tools used in the treatment of data included frequency counts, percentages, mean, cross-tabulation, one-way chi-square and t-test. Frequency counts, percentages, mean and chi-square were used to describe the respondent profile. Chi-square was also used to determine the gaming pattern of respondents. In this connection, 1% level of probability with 99% reliability was adopted to determine the degree of significance of the findings.

## RESULTS AND DISCUSSION

### Profile of the Respondents

Table 1 shows the profile of the respondents. Most of the respondents were from Manila (f = 368, 70%). It can also be noted that 30% (f = 155) of the respondents were non-Manila settlers, rather they were from the other parts of the National Capital Region (NCR) and the nearby provinces of Bulacan and Rizal. Most respondents were students (f = 382, 73%), had attained or were pursuing a college degree (f = 374, 72%), male (f = 356, 68%), young (19 and below) (f = 314, 60%), Roman Catholic (f = 423, 81%), single (f = 470, 90%), had a computer at home (f = 269, 51%), and belonged to the middle-income class (f = 334, 64%).

**Table 1 Profile of the Respondents**

| Profile | f | % | $\chi^2$ | *DF*[b] | *p*-value |
|---|---|---|---|---|---|
| **Place of residence** | | | | | |
| Manila settlers | 368 | 70 | – | – | – |
| Non-Manila settlers | 155 | 30 | | | |
| **Employement** | | | | | |
| Employed | 113 | 22 | | | |
| Non-employed | 28 | 5 | 391.78 | 2 | 0.000 |
| Student | 382 | 73 | | | |
| **Educational Background** | | | | | |
| Pre-school/Elementary | 29 | 6 | | | |
| High school | 106 | 20 | | | |
| Vocational | 10 | 2 | 930.82 | 4 | 0.000 |
| College | 374 | 72 | | | |
| Postgraduate | 4 | 1 | | | |
| **Sex** | | | | | |
| Male | 356 | 68 | 68.30 | 1 | 0.000 |
| Female | 167 | 32 | | | |
| **Age**[a] | | | | | |
| 19 and below | 314 | 60 | 21.08 | 1 | 0.000 |
| 20 and above | 209 | 40 | | | |
| **Religious Affiliation** | | | | | |
| Catholic | 423 | 81 | 199.48 | 1 | 0.000 |
| Non-Catholic | 100 | 19 | | | |
| **Civil Status** | | | | | |
| Single | 470 | 90 | | | |
| Married | 32 | 6 | 1,176.65 | 3 | 0.000 |
| Separated | 17 | 3 | | | |
| Widow/Widower | 4 | 1 | | | |
| **Computer Ownership** | | | | | |
| With computer at home | 269 | 51 | 0.43 | 1 | 0.512 |
| Without computer at home | 254 | 49 | | | |
| **Monthly Family Income** | | | | | |
| Lower-income class (P13,999 or $325 and below) | 189 | 36 | 50.20 | 1 | 0.000 |
| Middle-income class (P14,000-P56,000) ($326-$1,302) | 334 | 64 | | | |
| **TOTAL** | **523** | **100%** | | | |

[a] Average age is 19.80 years old
[b] *DF* – Degrees of freedom

Table 1 also showed that the respondent profile differed significantly except for home computer ownership. It was found out that most Internet users were students ($\chi 2$ = 391.78, $DF = 2$, $p < 0.01$). Adomi et al. (2004), Furuholt et al. (2005), Wahid et al. (2004), Ishii and Wu (2006), and Alao and Folorunsho (2007) also yielded similar findings. Moreover, most respondents were pursuing or had attained a college degree ($\chi 2$ = 930.82, $DF = 4$, $p < 0.01$). This was similar to the findings of Haseloff (2005), Adomi et al. (2004), and Adebayo (2008).

The male gender dominated internet usage in cyber cafés ($\chi 2$ = 68.30, $DF = 1$, $p < 0.01$). This finding supports the studies of Haseloff (2005), Furuholt et al. (2005), Adomi (2007), Alao and Folorunsho (2007), and Olatokun (2009). This study also revealed that most internet users were young – children and teenagers that belonged to the age group 19 and below ($\chi 2$ = 21.08, $DF = 1$, $p < 0.01$). Conversely, Furuholt et al. (2005), Adomi (2007), Alao and Folorunsho (2007), Adebayo (2008), and Olatokun (2009) found that most respondents were in the age group of 20–30 years old. In the Philippines, this age group would be categorized as young urban professionals.

Also similar to the findings of Furuholt et al. (2005), Alao and Folorunsho (2007), and Adebayo (2008), most of the respondents were single (never been married) ($\chi 2$ = 1,176.65, $DF = 3$, $p < 0.01$). Internet users also tended to be Roman Catholic ($\chi 2$ = 199.48, $DF = 1$, $p < 0.01$), and belong to the middle-income class ($\chi 2$ = 50.20, $DF = 1$, $p < 0.01$).

Finally, the number of respondents with computers at home and those without computers was not significantly different ($\chi 2$ = 0.43, $DF = 1$, $p < 0.01$). Interestingly, although more than half of the respondents had computers at home, they still availed the Internet services of a cyber café. Can this be attributed to the high cost of broadband Internet? Will the result of a future study in the Philippine setting support the study of Haseloff (2005)? Only future studies can answer these questions.

**Internet Usage and Pattern of Use**

Table 2 shows the Internet usage of respondents in terms of average hours spent per visit, frequency of visit within a week, and visiting time of day. On the average, respondents spent 3.30 hours per visit, tended to avail the Internet services of cyber cafés once or twice a week (f = 193, 36.9%), and usually during the afternoon (f = 274, 50.3%). A one-way chi-square showed that this finding was unlikely to have to arisen from sampling error (Once or twice a week, $\chi 2$ = 90.04, $DF = 3$, $p < 0.01$ and Afternoon, $\chi 2$ =

113.98, *DF* = 2, $p < 0.01$).

Meanwhile, Table 3 shows the average hours spent per visit categorized by frequency of visit within the week and visiting time of day. In general, findings revealed that as the frequency of visit increased, the average hour spent per visit of the Internet users also increased. As shown in Table 2, most of the respondents visited an Internet café in the afternoon. However, “afternoon” users were only on the Internet for an average of 2.87 hours. On the contrary, “morning” users spent more hours (mean = 5.43 hours) than did those who used the Internet at other times of the day.

This is an interesting finding because the morning can be considered as the period to start the day and yet respondents dedicate about 6 hours of Internet use in a public venue. It can be noted, however, that in this study “morning” referred to 12:01 am to 11:59 am. Survey distribution was conducted between 7 am to 11:59 am – roughly five hours. This could mean that respondents had been using the Internet before the survey was distributed, probably between 12:01 am to 11:59 am.

It was also found that most respondents were single (see Table 1). Does it mean “morning users” can dedicate their time in the morning since they do not yet have a commitment? Better yet, does civil status relate to morning Internet usage? This is a research question still unanswered.

**Table 2 Internet Usage**

| Cyber café usage | Frequency | Percent | $\chi^2$ | *DF* | *p*-value |
|---|---|---|---|---|---|
| **Frequency of visit** | | | | | |
| Once to twice a week | 193 | 36.9 | 90.04 | 3 | 0.000 |
| Three to four times a week | 149 | 28.5 | | | |
| Five to six times a week | 44 | 8.4 | | | |
| Every day | 137 | 26.2 | | | |
| **Visiting Time in a Day** | | | | | |
| Morning | 75 | 14.3 | 113.98 | 2 | 0.000 |
| Afternoon | 274 | 52.4 | | | |
| Evening | 174 | 33.3 | | | |
| **TOTAL** | **523** | **100%** | | | |
| | **mean** | | | | |
| **Hours spent per visit** | **3.30** | | | | |

Some Internet users spent, at most, 12 hours a week online. This may also imply that these were prudent Internet users as they judiciously spent fewer hours availing the Internet services of a cyber café.

It can also be noted that “five to six times a week” and “every day” users spent at least 19.75 hours a week online. Chou and Hsiao (2000) (as cited in Tahiroglu, Celik, Uzel, Ozcan, & Avci, 2008) reported that some respondents are problematic internet and users spent 20 to 25 hours a week online. Problematic internet use (PIU) is “a psychiatric condition involving maladaptive thoughts and pathological behaviors” (Tahiroglu et al., 2008) and it is characterized by “pervasive compulsion to be online” (Davis, Flett, & Besser, 2002).

This study, however, did not identify whether respondents were PIUs or not. Further, no studies have yet determined whether cyber cafes’ Internet users are PIUs or not, nor has any study examined the Internet activities of these users in the Philippines. In other words, little is known about these users. As such, numerous studies can be initiated to make (1) comparison of the characteristics of prudent and problematic internet users, (2) determine the purposes and services they avail in cyber cafés, (3) asses their online behaviors, and (4) examine the effects or impacts of the Internet usage of these two types of users.

**Table 3 Cross-Tabulation of Frequency of Visit and Visiting Time in a Day by Average Hour Spent Per Visit**

| Cyber cafe usage | Average hour spent per visit | Average hour spent per week (Average hours spent per visit × frequency of visit) |
|---|---|---|
| **Frequency of visit** | | |
| Once to twice a week | 2.13 | 2.13 – 4.26 |
| Three to four times a week | 3.05 | 9.15 – 12.2 |
| Five to six times a week | 3.95 | 19.75 – 23.7 |
| Every day | 4.99 | 34.93 |
| **Visiting Time in a Day** | | |
| Morning | 5.43 | – |
| Afternoon | 2.87 | – |
| Evening | 3.08 | – |

Further analysis also showed that when the profile of the respondents was cross-tabulated with frequency of visit (see Table 4) to avail Internet services in a cyber café, it was shown that respondents tended to visit cyber cafés once or twice a week. This was finding was significant throughout the profile of the respondents (Student, $\chi 2$ = 46.08, *DF* = 3, $p < 0.01$; College, $\chi 2$ = 80.03, *DF* = 3, $p < 0.01$; Male, $\chi 2$ = 44.52, *DF* = 3, $p < 0.01$; 19 and below, $\chi 2$ = 35.40, *DF* = 3, $p < 0.01$; Catholic, $\chi 2$ = 82.20, *DF* = 3, $p < 0.01$; Single, $\chi 2$ = 71.74, *DF* = 3, $p < 0.01$; Middle-income class, $\chi 2$ = 59.04, *DF* = 3, $p < 0.01$). This finding suggests that, regardless of the demographic characteristics (i.e., profile of the respondents) of respondents, they tended to visit cyber cafés once or twice a week.

**Table 4 Profile of the Respondents Cross-Tabulated With Frequency of Visit**

| Profile | Internet Usage in Cyber Café by Frequency of Visit | | | | Total | $\chi^{2c}$ | *p*-value |
|---|---|---|---|---|---|---|---|
| | Once or twice a week | Three or four times a week | Five of six times a week | Every day | | | |
| Employment – Student | 127 | 112 | 40 | 103 | 382 | 46.08 | 0.000 |
| Educational Background – College | 148 | 108 | 28 | 90 | 374 | 80.03 | 0.000 |
| Sex – Male | 114 | 104 | 35 | 103 | 356 | 44.52 | 0.000 |
| Age – 19 and below | 97 | 91 | 33 | 93 | 314 | 35.40 | 0.000 |
| Religious Affiliation – Catholic | 164 | 118 | 34 | 107 | 423 | 82.20 | 0.000 |
| Civil Status – Single (Never been married) | 162 | 140 | 41 | 127 | 470 | 71.74 | 0.000 |
| Monthly Family Income – Middle-income class | 127 | 99 | 31 | 77 | 334 | 59.04 | 0.000 |

[c]*DF = 3*

Meanwhile, it is also found that respondents tend to visit cyber cafés in the afternoon regardless of their demographics (Students, $\chi 2$ = 71.74, *DF* = 3, $p < 0.01$; College, $\chi 2$ = 71.74, *DF* = 3, $p < 0.01$; Male, $\chi 2$ = 71.74, *DF* = 3, $p < 0.01$; Young (19 years old and below), $\chi 2 = 71.74$, $DF = 3$, $p < 0.01$; Roman Catholic, $\chi 2 = 71.74$, $DF = 3$, $p < 0.01$; Single, $\chi 2 = 71.74$, $DF = 3$, $p < 0.01$; and Middle-income class, $\chi 2 = 71.74$, $DF = 3$, $p < 0.01$). This finding also identified another research gap; the reason of such visit in the afternoon must be investigated.

**Table 5 Profile of the Respondents Cross-Tabulated with Frequency of Visiting Time in a Day**

| Profile | Internet Usage in Cyber Café by Frequency of Visit | | | Total | $\chi^{2c}$ | *p*-value |
|---|---|---|---|---|---|---|
| | Morning | Afternoon | Evening | | | |
| Employment – Student | 50 | 213 | 119 | 382 | 105.15 | 0.000 |
| Educational Background – College | 46 | 197 | 131 | 374 | 91.93 | 0.000 |
| Sex – Male | 50 | 189 | 117 | 356 | 81.44 | 0.000 |
| Age – 19 and below | 39 | 177 | 98 | 314 | 91.61 | 0.000 |
| Religious Affiliation – Catholic | 60 | 224 | 139 | 423 | 95.42 | 0.000 |
| Civil Status – Single (Never been married) | 63 | 249 | 158 | 470 | 110.43 | 0.000 |
| Monthly Family Income – Middle-income class | 55 | 170 | 109 | 334 | 59.47 | 0.000 |

[c] *DF = 2*

## SUMMARY, CONCLUSIONS AND RECOMMENDATIONS

In summary, it has been shown that the profiles of Internet users vary. Overall, respondents spend 3.30 hours per visit and tend to visit cyber cafés one to two times a week in the afternoon. The average hour spent per visit by "five to six times a week" and "every day" users was at least 19.75 hours a week. Internet users spent longer hours in the morning than any other time of day. When categorized by demographics of the respondents, it was consistently found out that respondents tended to visit cyber cafés one to two time a week, usually in the afternoon.

The findings show that frequency of visit of the respondents was not equally distributed throughout the week when grouped according to profile. It was also revealed that respondents showed a tendency to use the Internet in a cyber café at a particular time of the day when grouped according to profile. Thus, the first hypothesis was accepted while the second hypothesis was rejected. In other words, respondents tended to use the Internet in cyber cafés once or twice a week every afternoon.

It was also found that there are two types of users. The first type includes those who use the Internet more than 20 hours a week. It is possible that these users are problematic Internet users. The second type consists of respondents who use the Internet judiciously and are referred to as prudent Internet users.

This study also recognizes limitations and also open avenues for future research. First, the profile of Internet users in terms of employment, education, sex, age, religious affiliation, civil status, and economic status varies. An in-depth study can be initiated to determine the existence of these variations. For instance, researchers can investigate the dominance of males on Internet usage, the influence of broadband costs for home Internet access, and the relationship of civil status to morning Internet usage.

Second, although it was shown that most of respondents tended to visit an internet café in the afternoon, the reasons for such visits in the afternoon are not clear. Similarly, researchers can also investigate morning Internet browsing of the respondents. The study of Adomi (2007) can serve as basis for such a study.

Third, although it is not the intention of the current study to determine the purpose of and the services availed by respondents, there is a need for such study. A study of Internet usage per respondent profile can also be initiated (e.g., students' internet usage, college students' internet usage, males' internet usage, etc.).

Lastly, a study can be initiated as to whether cyber café Internet user are PIUs, with the determination of the online activities of prudent and problematic internet users, respectively. Further, users online behaviors, the purposes, and services they avail in cyber cafés, and the impact or effects of such usage can also be studied.

The social implications of publicly accessed Internet must be studied because it imposed serious threats to its users – either problematic users or as internet addicts (Yellowlees & Marks, 2007). This paper seeks collaboration of local government units for strict implementation of controlling and monitoring cyber café business activities and cyber café owners for a sense of business social responsibility. Thus, this study calls for a multi-disciplinary approach to address these gaps.

## REFERENCES

Adebayo, O.S. (2008). Performance evaluation and indices of cyber café business: A factor analytic approach. *Journal of Information and Communication Technology, 7*, 89-102.

Adomi, E.E. (2007). Overnight Internet browsing among cybercafé users in Abraka, Nigeria. *The Journal of Community Informatics, 3*(2). Retrieved October 12, 2010, from http://ci-journal.net/index.php/ciej/article/view/322/351

Adomi, E.E., Okiy, R.B., & Ruteyan, J.O. (2003). A survey of cybercafés in Delta State, Nigeria. *The Electronic Library, 21*(5), 487-495. doi:10.1108/02640470310499876.

Adomi, E.E., Omodeko, F.S., & Otolo, P.U. (2004). The use of cybercafé at Delta State University, Abraka, Nigeria. *Library Hi Tech, 22*(4), 383-388. doi:10.1108/07378830410570485.

Alao, I.A., & Folorunsho, A.L. (2007). The use of cybercafés in Ilorin, Nigeria. *The Electronic Library, 26*(2), 238-248. doi:10.1108/02640470810864127.

Chou, C., & Hsiao, M.C. (2000). Internet addiction, usage, gratification and pleasure experience: The Taiwan college students' case. *Computers & Education, 35*(1), 65-80. doi:10.1016/S0360-1315(00)00019-1.

Davis, R.A., Flett, G.L., & Besser, A. (2002). Validation of a new scale for measuring problematic Internet use: Implications for preemployment screening. *CyberPsychology & Behavior, 5*(4), 331-345. doi:10.1089/109493102760275581.

Furuholt, B., Wahid, F., & Kristiansen, S. (2005). Information dissemination in a developing society: Internet café users in Indonesia. *The Electronic Journal of Information Systems in Developing Countries, 22*(3), 1-16.

Globalization and World Cities (GaWC) Study Group and Network (2008). *The world according to GaWC 2008*. Retrieved October 25, 2010, from http://www.lboro.ac.uk/gawc/world2008t.html

Griffiths, M.D., Davies, M.N.O., & Chappell, D. (2004). Demographic factors and playing variables in online computer gaming. *CyberPsychology & Behavior, 7*(4), 479-487. doi:10.1089/cpb.2004.7.479.

Gurol, M., & Sevindik, T. (2006). Profile of internet café users in Turkey. *Telematics and Informatics, 24*(1), 59-68. doi:10.1016/j.tele.2005.12.004.

Haseloff, A.M. (2005). Cybercafes and their potential as community development tools in India. *The Journal of Community Informatics, 1*(3), 53-65.

Ishii, K., & Wu, C.-I. (2006). A comparative study of media cultures among Taiwanese and Japanese youth. *Telematics and Informatics, 23*(2), 95-116. doi:10.1016/j.tele.2005.05.002.

National Statistics Office (2008). *National Statistics 2007 census of population*. Retrieved November 8, 2010, from http://www.census.gov.ph/data/census2007/index.html

Olatokun, W.M. (2009). Analysing socio-demographic differences in access and use of ICTs in Nigeria using the capability approach. *Issues in Informing Science and Information Technology, 6*, 479-496.

Tahiroglu, A.Y., Celik, G.G., Uzel, M., Ozcan, N., & Avci, A. (2008). Internet use among Turkish adolescents. *CyberPsychology & Behavior, 11*(5), 537-543. doi:10.1089/cpb.2007.0165.

Wahid, F., Furuholt, B., & Kristiansen, S. (2004). Global diffusion of the internet III: Information diffusion agents and the spread of internet cafés in Indonesia. *Communications of the Association for Information Systems, 13*(1), 589-614.

Williams, D., Yee, N., & Caplan, S.E. (2008). Who plays, how much, and why? Debunking the stereotypical gamer profile. *Journal of Computer-Mediated Communication, 13*(4), 993-1018. doi:10.1111/j.1083-6101.2008.00428.x.

Yellowlees, P.M., & Marks, S. (2007). Problematic internet use or internet addiction? *Computers in Human Behavior, 23*(3), 1447-1453. doi:10.1016/j.chb.2005.05.004.